\documentclass[fleqn,usenatbib]{mnras}

\usepackage{newtxtext,newtxmath}

\usepackage[T1]{fontenc}
\usepackage{ae,aecompl}
\usepackage{deluxetable}

\usepackage{graphicx}	
\usepackage{amsmath}	
\usepackage{amssymb}	

\makeatletter
\def\@to{to}
\makeatother

\title[Long term variability of Swift J1753.5-0127]{Long term variability of Swift J1753.5-0127: \\ X-ray spectral-temporal correlations during state transitions}

\author[Qingcui Bu et al.]{
Qingcui Bu,$^{1}$\thanks{E-mail: buqc@ihep.ac.cn}
Lian Tao,$^{1}$
Yu Lu,$^{1}$
Shuangnan Zhang,$^{1}$
Liang Zhang,$^{2}$
Yue Huang,$^{1}$ \\
\newauthor
Li Chen,$^{3}$ 
Jinlu Qu,$^{1}$
and Xiang Ma$^{1}$
\\
$^{1}$Key Laboratory of Particle Astrophysics, Institute of High Energy Physics, CAS, Beijing 100049, China\\
$^{2}$School of Physics and Astronomy, University of Southampton, Southampton, SO17 1BJ, UK\\
$^{3}$Department of Astronomy, Beijing Normal University, Beijing 100875, China\\
}

\date{Accepted XXX. Received YYY}

\pubyear{2019}

\begin{document}
\label{firstpage}
\pagerange{\pageref{firstpage}--\pageref{lastpage}}
\maketitle

\begin{abstract}
	
We studied the long-term evolution of the spectral-temporal correlated properties of the black hole candidate Swift J1753.5-0127 from the onset of its outburst until 2011 with the Rossi X-ray Timing Explorer (\textit{RXTE}). The source stayed most of its lifetime during hard state, with occasionally transitioned to the hard intermediate state. Similar to typical black hole transients, Swift J1753.5-0127 traces a clear hard line in absolute rms - intensity diagram during the low hard state, with expected highest absolute rms, while shows a clear turn during the hard intermediate state, accompanied by lower absolute rms. Different from Cyg X-1, we found that frequency-dependent time lag increased significantly in the 0.02 -- 3.2 Hz band during state transition in this source. The X-ray time lags in 0.02 -- 3.2 Hz can therefore be used as indicators of state transition in this source. Type-C QPO frequency is positively related with its fractional rms and X-ray photon index, suggesting a moving inwards disk/corona scenario. We discussed the physical interpretation of our results in our paper. 
\end{abstract}



\section{Introduction} \label{sec:intro}

Black hole low mass X-ray binaries (BH-LMXB) mostly are transient systems in which a black hole accretes matters from its companion star via an accretion disc. Black hole transients (BHTs) spend most of their lifetime in quiescent state at a luminosity below $10^{33} \rm{erg/s}$, and show occasional outbursts that lasts from weeks to months. During an outburst, the source luminosity can reach the Eddington limit, and the energy spectral properties change dramatically, as does the fast variability.

Low frequency quasi-periodic oscillations (LFQPOs) are commonly observed in BHTs, which are defined as narrow, but not coherent, peaks in the Fourier power density spectrum (PDS) computed from a corresponding lightcurve. Their frequencies typically varying from few mHz to tens of hertz. In BH systems, LFQPOs are known as three types: Type-A, B, and C \citep{cas05, mot15}. type-C QPOs are the most frequently observed type, and tightly related to the state evolution \citep{bel16}.

Despite that the QPOs have been extensively studied for decades, their origin is under debate. There are several models proposed to investigate the physical mechanism behind type-C QPOs: they generally consider either instability in the accretion flow \citep{tag99, tit99} or an effect of relativistic precession within the accretion flow \citep{ste99, ing09}. The orbital inclination dependence of LFQPOs found in recent work suggested that the type-C QPOs have a geometric origin \citep{mot15, van16}, which makes the Lense-Thirring (LT) precession model the most promising model in explaining the origin of type-C QPOs. \citet{ing09} proposed an updated truncated disk model based on LT precession to explain both type-C QPOs and their associated noise. In this model, the type-C QPOs are produced from the LT precession of the inner hot flow, while the associated noise arises from the variations in the mass accretion rate. However, the origin of type-B and type-A QPOs is much less clear. No comprehensive model has been proposed for type-B and type-A QPOs.

A clear pattern of X-ray spectral evolution of BHTs is found in most systems when a hardness-intensity diagram (HID) is applied, which is known as the q-shaped HID \citep{hom01}. During an outburst, BHTs start with low hard state (LHS) at low intensity and stay in LHS over a wide range of luminosity before it turning into the hard intermediate state (HIMS). During the LHS, the X-ray spectrum is dominated by a hard power law component with a photon index of $\sim$ 1.5 -- 1.8, which is believed to originate from an optical thin corona \citep{hom01, don07}. Type-C quasi-periodic oscillations (QPOs) with strong broad-band noise are usually observed in LHS and HIMS and their frequency often increase with source luminosity. The transition to soft intermediate state (SIMS) from HIMS happens when the spectra become softer. The root-mean-square (rms) variability amplitude in the SIMS is weaker compared to the HIMS, with the appearances of type-A/B QPOs in the corresponding power spectra. Type-A/B QPOs are often accompanied by weak red noise, thereby makes them distinguishable from type-C QPOs \citep{wij99, rem02, cas05}. The source turns into the high soft state (HSS) when the disc component becomes dominant in the energy spectra. With the decrease of luminosity, the source turns into LHS again after passing through the intermediate states in reverse. However, in some cases, BHTs spend the whole outburst in the LHS/HIMS and never transit to the HSS/SIMS \citep{cap09}, which is generally referred as a ``failed" outburst. For instance, the 2010 outburst of MAXI J1659-152 \citep{deb15}, the 2011 outburst of MAXI J1836-194 \citep{jan16}, and the 2000 outburst of XTE J1118+480 \citep{cha16}.

The relation between X-ray fast variability and spectral state evolution in BHTs has been addressed in many work, hence makes fast variability an indicator to trace the source state during the outburst. Considerable work has been done to study the state evolution of BHTs, suggesting that the rms-intensity diagram (RID) provides a common framework to describe accretion states in BHTs \citep{mun14, mot17}. The HSS usually occurs at relatively high luminosity and is dominated by a thermal component, while only weak/absent variability is seen in the HSS. Compared to HSS, the disk/corona geometry in LHS is less certain. All the important physical components can be observed in the LHS, including accretion disk, hot plasma (a corona or hot inner flow) and outflow, thereby makes it possible to improve current accretion models by studying long term spectral-temporal evolution in LHS. Prominent fast variability is commonly observed in the LHS of BHTs, which is generally combined of several noise "breaks" and occasionally presence of QPOs. 

Swift J1753.5-0127 was discovered by Swift/Burst Alert Telescope (BAT) on 2005 May 30 \citep{pal05} and had been active since then before going into quiescence state in 2016 November. After its discovery, the source spent most of its outburst time in LHS, and occasionally evolves into HIMS \citep{yos15}. However, the source did not transit in to HSS like most BHTs and returned to LHS instead \citep{sol13}. No signatures of soft states are found until 2015 March \citep{sha16}, with the subsequent follow-up observations from Swift X-ray Telescope (XRT). Motivated by this unique feature, in this work, we focus on studying the long term variability properties of Swift J1753.5-0127 during state transitions basing on the power density spectra (PDS), the X-ray time lags, and energy spectra analysis. 

This paper is structured as follows: Observations and data analysis are given in Section \ref{sec:data}, and results are present in Section \ref{sec:results}. Detailed discussion is arranged in Section \ref{sec:dis}. We briefly summarize our conclusions in Section \ref{sec:con}.  


\section{Observations and Data Analysis} \label{sec:data}

\subsection{Light curve and Hardness-Intensity Diagram}

We analyzed all the archival Rossi X-ray Timing Explorer (\textit{RXTE}) observations of Swift J1753.5-0127 from the onset of its 2005 outburst until November 29, 2011, for a total of 344 observations. For all the observations, energy spectra and light curves were extracted by applying a standard criteria: elevation angle $> 10^{\circ}$, pointing offset $< 0.01$ and South Atlantic Anomaly exclusion times of 30 minutes, under {\sc HEASOFT} v6.24.

\begin{figure}
	\centering
	\includegraphics[scale=.55]{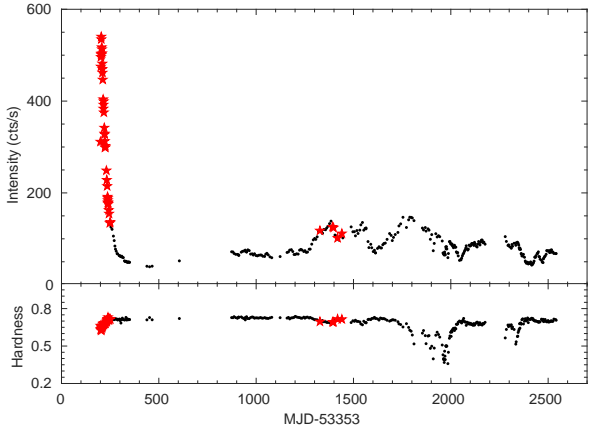}
	\caption{X-ray light curve (up panel) and hardness (bottom panel) evolution of Swift J1753.5-0127, from the onset of its 2005 outburst until November 29, 2011. Each point represents one PCA observation. Intensity and hardness are defined in Section 2.2. The red filled stars represent QPO observations. \label{fig:1}}
\end{figure}

\begin{figure}
	\centering
	\includegraphics[scale=.5]{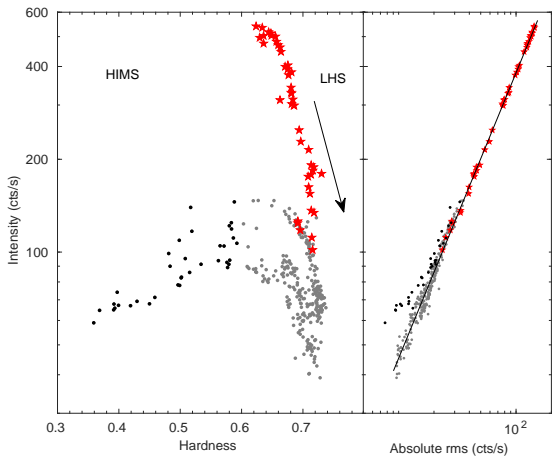}
	\caption{The hardness - intensity (left) and absolute rms - intensity (right) diagrams of Swift J1753.5-0127. Each point represents one PCA observation. The absolute rms was calculated in 0.02 -- 32 Hz frequency band. The red filled stars mark the QPO observations and the black arrow indicates the evolution path. The low hard state (grey dot) and hard intermediate state (black dot) are distinguished by dash line in the figure. \label{fig:2}}
\end{figure}

We used the overall 344 PCA observations to produce the Standard 2 (STD2) light curve and the hardness in Fig.　\ref{fig:1}, with background dead-time correction. The light curve was computed from the STD2 channel 0 -- 31 ( $\sim$ 2 -- 15 keV) and the spectral hardness ratio was defined as the ratio of count rates in the STD2 channel bands 11 -- 20 ($\sim$ 6 -- 10 keV) and 4 -- 10 ($\sim$ 3 -- 6 keV). We used only PCU 2 data to calculate the count rate and colors.\footnote{The response function variation from one observation to another is ignored here, since the light curves were created in approximately energy bands.}In Fig.　\ref{fig:2}, we plot the HID and RID of Swift J1753.5-0127, with the total fractional rms amplitude averaged in 0.02 -- 32 Hz frequency range. QPO observations are marked with red stars in Fig.　\ref{fig:1} and Fig. \ref{fig:2}.

\subsection{Timing analysis} 

For each observation, power density spectrum (PDS) was averaged from 32s-long stretches from the high time resolution data mode in the energy range $\sim$ 2 -- 15 keV (PCA absolute channel 0-35), with a Nyquist frequency of 1024 Hz. We used the {\sc GHATS} \footnote{http://www.brera.inaf.it/utenti/belloni/GHATS Package/Home.html} package under {\sc IDL} to create the 32s-long stretches, then averaged them together. The contribution of the Poissonian noise was subtracted according to \citet{zha95}. The PDS were further converted into square fractional rms \citep{bel90} after applying Leahy normalization \citep{lea83}. 

\begin{figure}
	\centering
	\includegraphics[scale=.3]{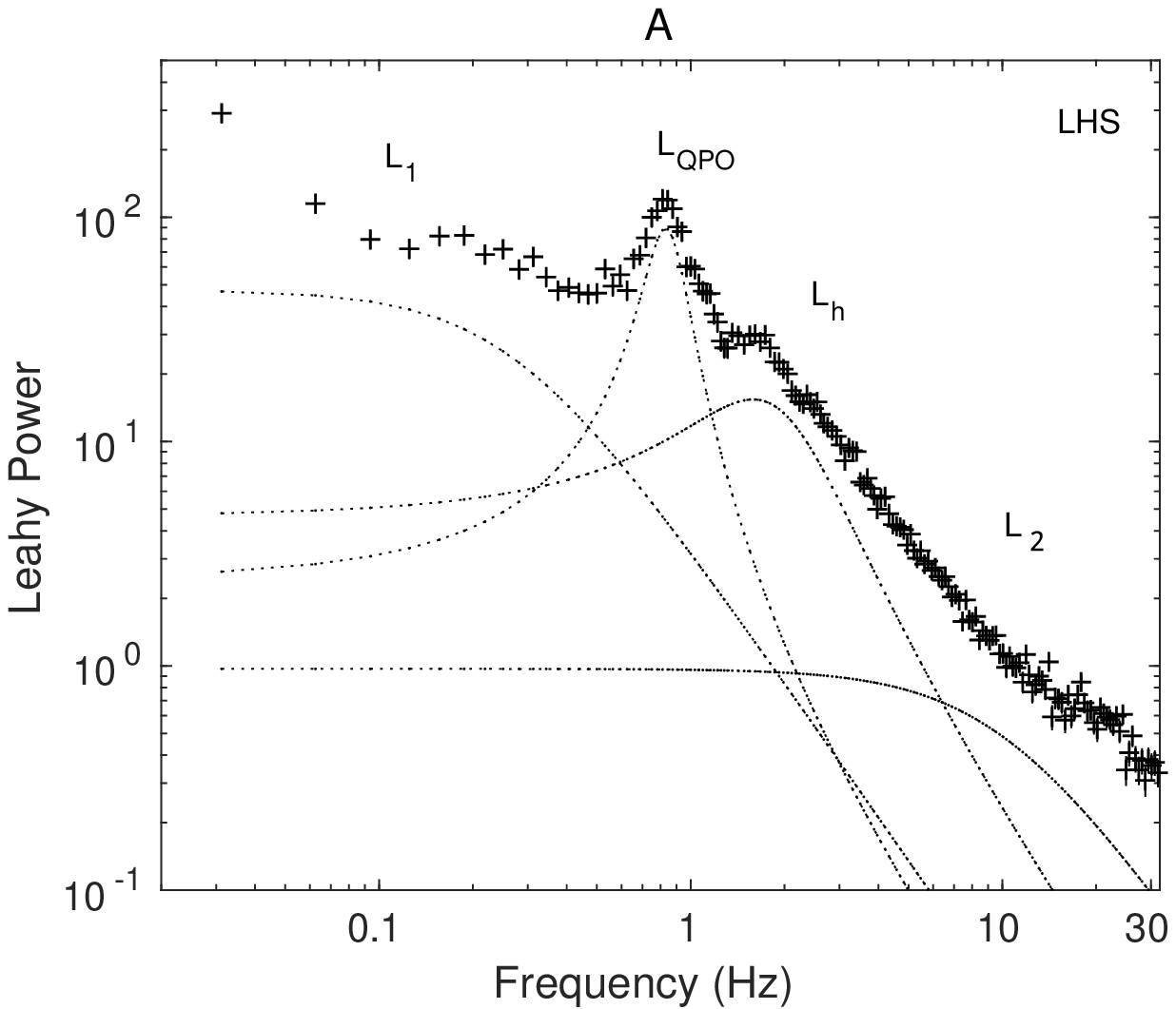}
	\includegraphics[scale=.3]{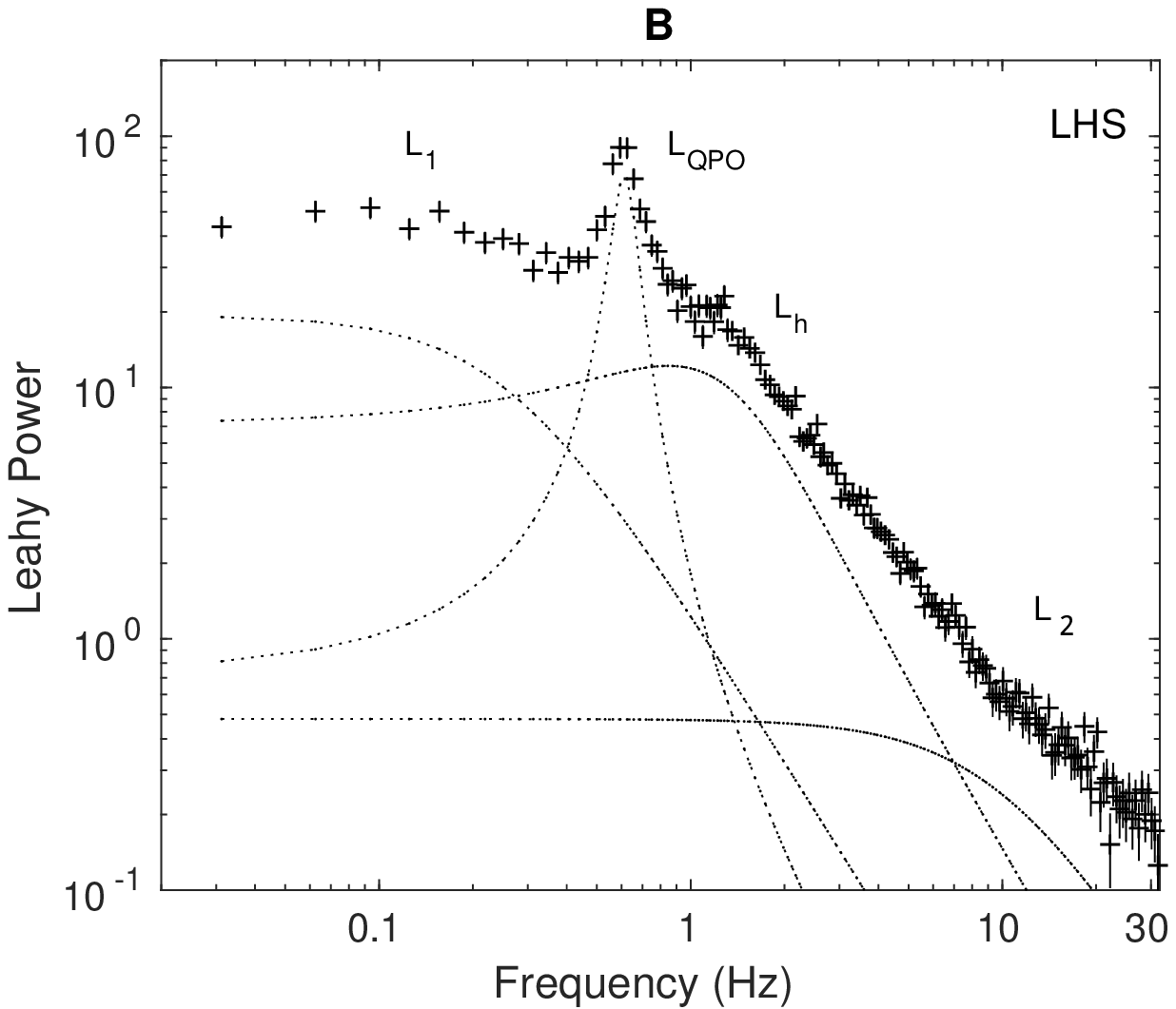}\\
	\includegraphics[scale=.3]{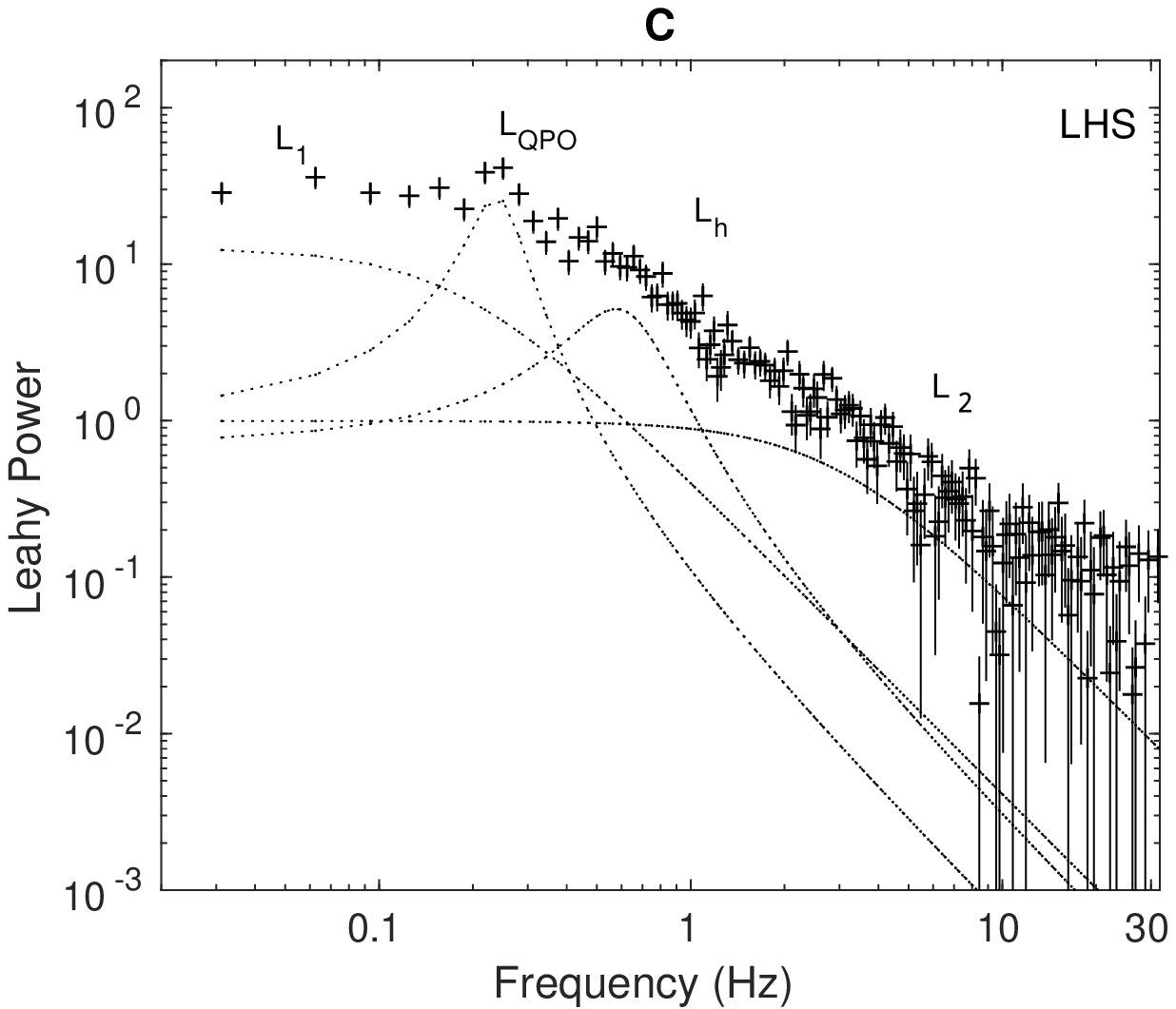}
	\includegraphics[scale=.3]{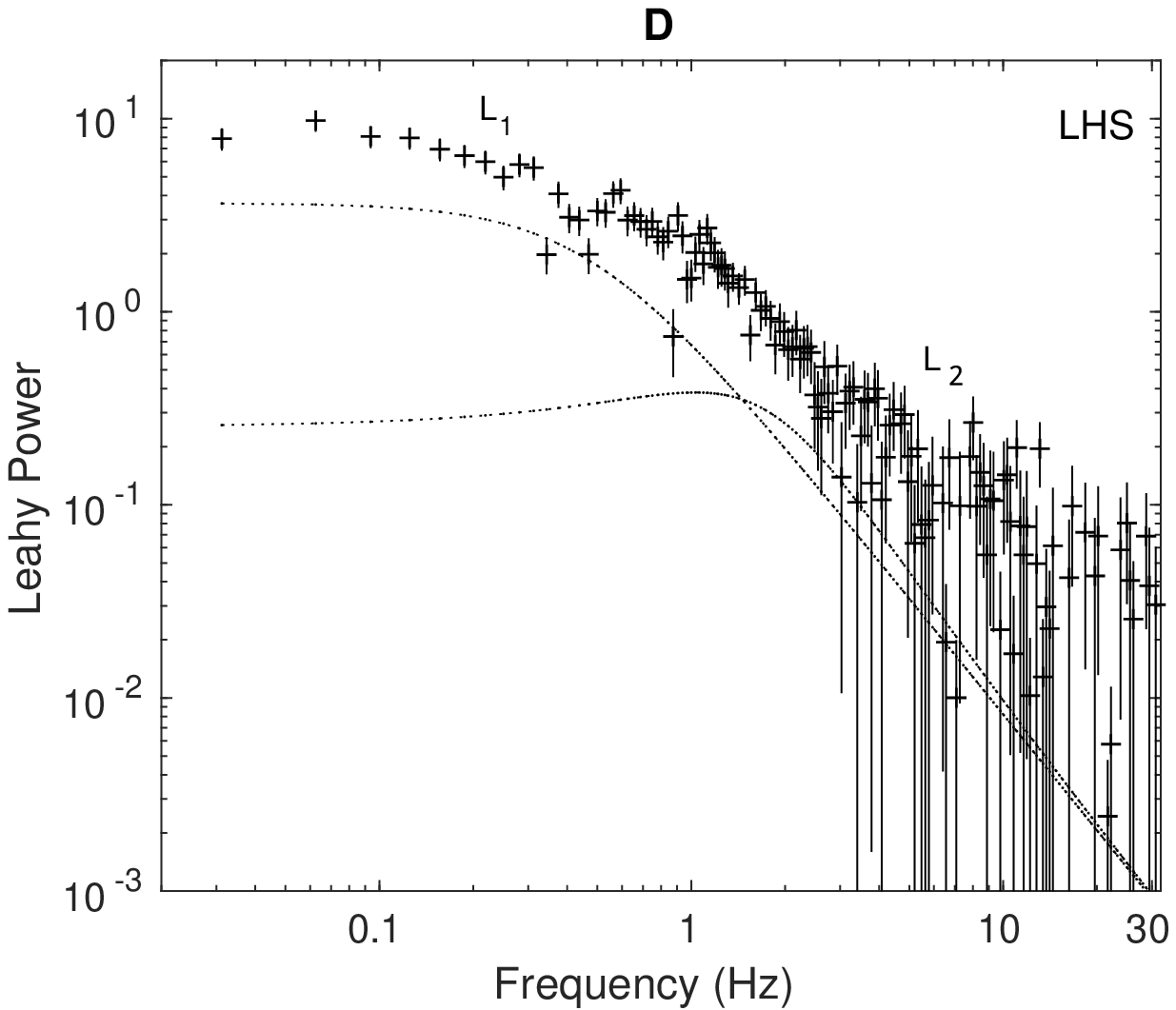}\\
	\includegraphics[scale=.3]{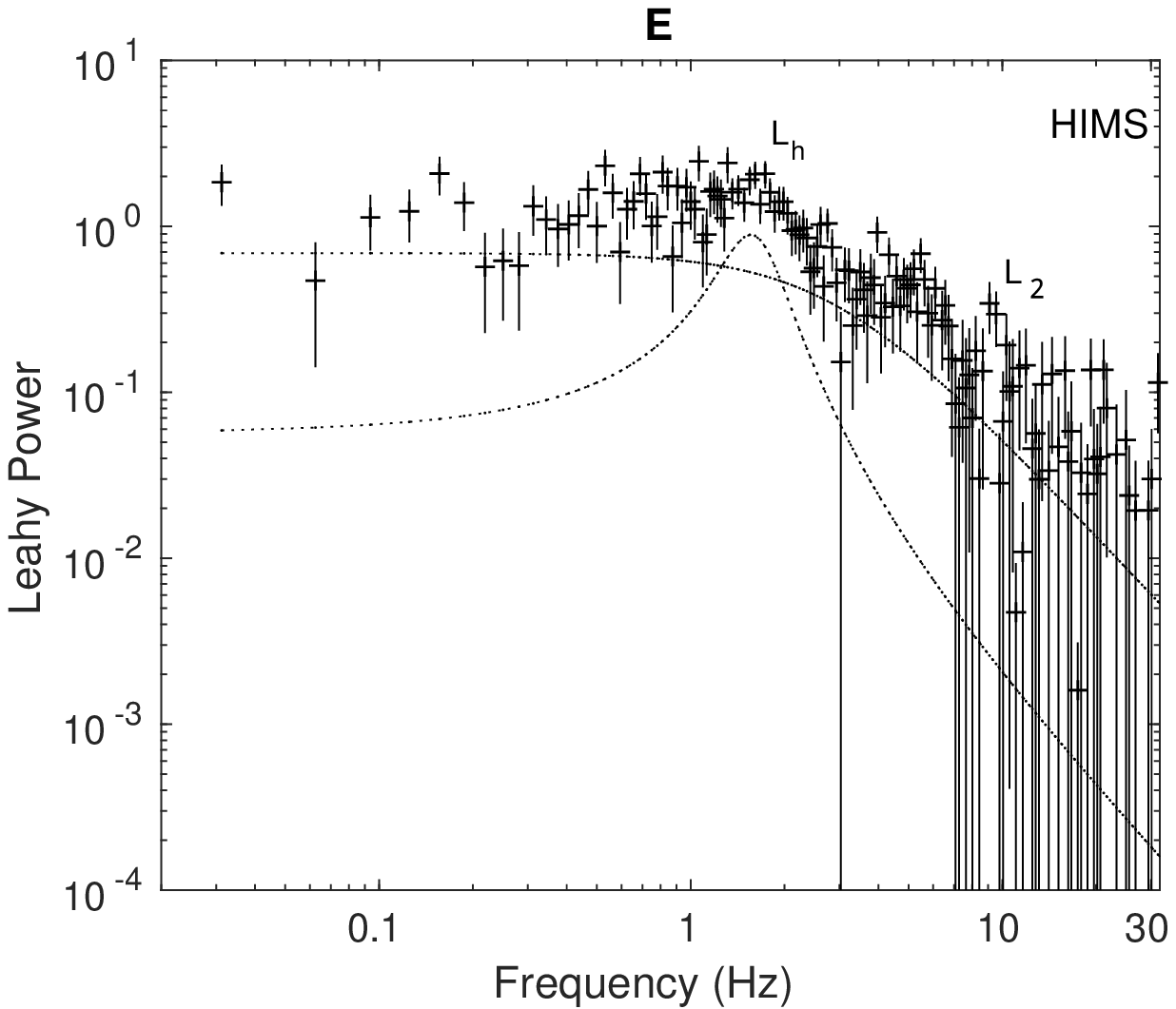}
	\includegraphics[scale=.3]{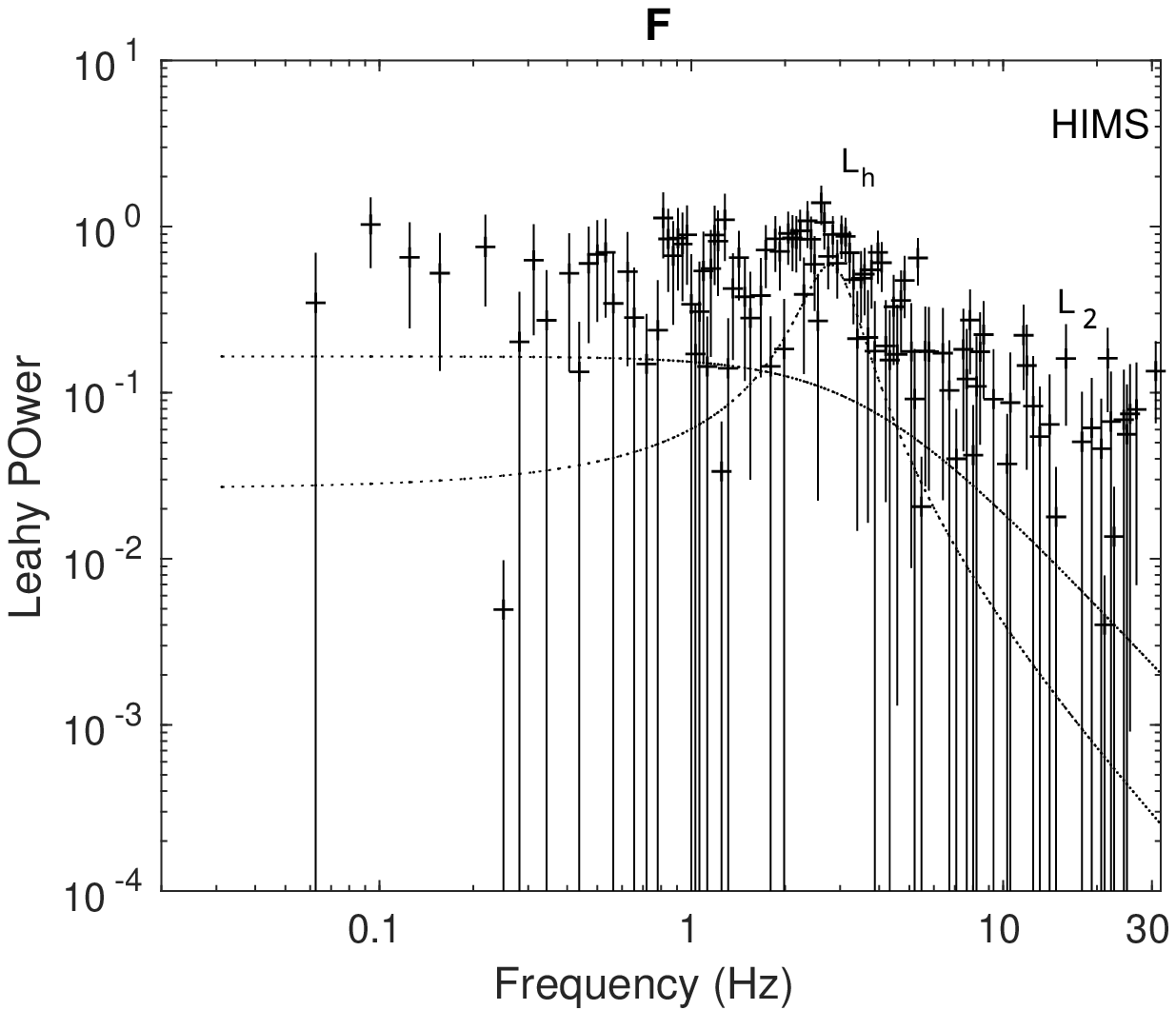}
	\caption{Fits to six representative PDS of Swift J1753.5-0127. The QPO ($L_{QPO}$), hump ($L_h$), noise components ($L_1$ and $L_2$) and the source state are labeled in the plots, together with the fitting functions. Figure A-F are arranged according to the temporal sequence, corresponding to ObsID 91094-01-01-04, 91423-01-02-00, 91423-01-06-04, 93105-02-02-00, 95105-01-15-00 and 95105-01-13-03 separately. \label{fig:3}}
\end{figure}

PDS fittings were taken under the {\sc xspec} (version 12.10.0c), by applying a one-to-one energy-frequency conversion with a unit response. We used a multi-Lorentzian function model to fit all the PDS. In particular, the broad band noises are fitted with two broad zero-centered Lorentzians ($L_1$ and $L_2$), occasionally low frequency humps ($L_h$) are fitted with broad Lorentzian and the QPOs ($L_{\rm{QPO}}$) are fitted with narrow Lorentzian. Errors on the fit parameters were determined at $1\sigma$ confidence level. Examples of six PDS from different source states are shown in Fig. \ref{fig:3}.


In this work, we focus our temporal variability study on the low frequency band. Hence, the total integrated rms was averaged in the frequency band 1/32 -- 32 Hz. For QPO study, only peaks with significance $>3\sigma$ were selected for further analysis. 

Time lags are widely used techniques in analyzing signals detected in different energy bands, with the application of the cross-spectra to measure the delays at different timescales. The average X-ray time lag in the 3.2 -- 10 Hz range has drawn attention in Cyg X-1, which is thought to be related with state transitions \citep{pot00, pot03}. Swift J1753.5-0127 is found being similar to Cyg X-1 in several ways, for instances, both sources experienced frequent spectral softenings and hardenings during outburst, both failed to detect type-C QPOs during hard intermediate states, and followed similar spectral parameters correlations \citep{pot03, sol13}. Motivated by their work, we computed two concurrent light curves from a low-energy ($\sim$ 2 -- 6 keV) and a high-energy channels ($\sim$ 6 -- 15 keV), and averaged their time lag in the 3.2 -- 10 Hz range. To make a further comparison with other frequency bands, we measured the time lags in the 0.02 -- 3.2 Hz, 10 -- 20 Hz and 0.02 -- 32 Hz bands as well. The QPO time lag is averaged in $\nu_{QPO} \pm \rm{FWHM}/2$, $\nu_{\rm{QPO}}$ is the QPO centroid frequency and the FWHM is its full width at half maximum.

\subsection{Spectral analysis}

For spectra analysis, the \textit{RXTE}/PCU 2 (3.0 -- 25 keV) and HEXTE-B (20 -- 200 keV) energy spectra were combined through a scaling factor. Background and dead-time corrections were applied separately to the two instruments. All the spectra were extracted with the {\sc HEASOFT} version 6.24. When selecting HEXTE data, only observations taken from the instruments actively rocking period were considered. Since cluster A failed to rock since 2006 October. Only data from cluster B was included in this work, thereby makes a total of 189 observations by the time cluster B stopped rocking in 2009 December.

All spectra fits were carried out using the {\sc xspec} 12.10.0c package, in which the best-fitting model is evaluated using the minimum $\chi^2$ techniques. The interstellar absorption (\textit{tbabs}) was fixed to the averaged column density value ($N_{\rm{H}}=2.0\times10^{21} \rm{cm^{-2}}$), as suggested by \citet{fro14}. A systematic error of 0.6$\%$ was added, given the uncertainties in the instrument calibration. 

The energy spectra of BHTs are generally described by a combination of blackbody and power law components. To find the proper components for the energy spectra, we performed several trails. When applied spectra fitting during the hard state, we modeled the hard spectra with a simple absorbed broken powerlaw (\textit{bknpow}), or a \textit{cutoffpl}, or a \textit{powerlaw} model. However, the break point in \textit{bknpow} model was constantly insensitive to the model and sometimes a \textit{powerlaw} model can not fit the hard component. Therefore, a \textit{cutoffpl} model was introduced to fit the hard spectra. When the e-folding energy cutoff can not be well constrained, we fix the cutoff energy at 1 MeV\footnote{In this case, a \textit{cutoffpl} model can be used as a \textit{powerlaw} model}. An additional disk blackbody (\textit{diskbb}) and sometimes a gaussian emission line (\textit{gaussian}) were required at low energies in addition to \textit{cutoffpl}. A \textit{diskbb} component was included in all the fits to get consistent fitting results. In the cases of whether to add a gaussion line, we used an F-test to have an indication of the improvement of the fits \citep{pro02}. Errors on the fit parameters were determined at $1\sigma$ confidence level.

\section{Results} \label{sec:results}

In this section, we first study the long term evolution of Swift J1753.5-0127 basing on X-ray light curve and HID (Section 3.1), then we discuss the temporal-spectral correlated properties basing on fitting parameters, i.e., the X-ray time lags and the photon index correlations during the state transitions (Section 3.2).

\subsection{Outburst evolution}

In Fig. \ref{fig:1}, the overall light curve shows a fast rise (first two observations) and exponential decay, corresponding to an increase in the hardness ratio (from 0.62 to 0.72), suggesting that the source is undergoing the LHS. During this time, a number of type-C QPOs (red stars) was observed, characterized with a strong associated flat-top noise in the corresponding PDS. 

\begin{figure*}
	\includegraphics[scale=.45]{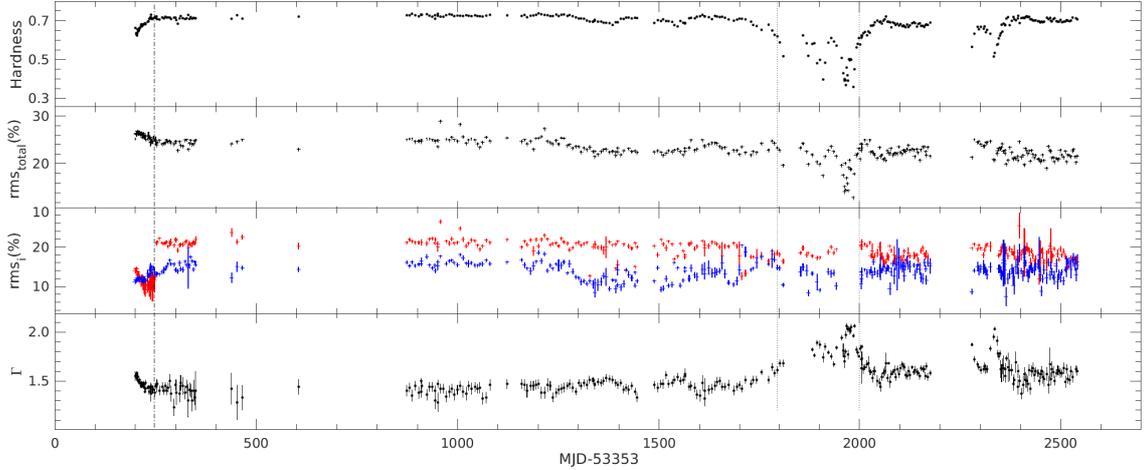}
	\caption{The hardness, total fractional rms ($rms_{\rm{total}}$), noise fractional rms (rms$_1$ and rms$_2$)and photon index ($\Gamma$) as a function of time (from the top to bottom). Each point represents one observation. The red crosses in the third panel represent the fractional rms of the low frequency noise component ($L_1$) and the blue ones represent the high frequency noise ($L_2$). The HIMS interval and the beginning part of the outburst are highlighted by dash lines. \label{fig:4}}
\end{figure*}

The source follows a hard line in the RID from the onset of the outburst, with the absolute rms decreasing from 142 cts/s to $\sim$ 100 cts/s (fractional rms from $\sim 27\%$ to $\sim 22\%$ in Fig. \ref{fig:4}). The source keeps hardening until MJD 54600, after which the spectrum gradually softens for several times. A apparent soften occurs near MJD 55338, where the source reaches its lowest hardness (hardness $\sim 0.35$, see Fig. \ref{fig:2}), corresponding to a highest photon index ($\Gamma \sim 2.1$) and a lowest fractional rms ($\sim 12\%$). However, the fractional rms is significantly high compared to the soft state ( a few percent), revealing that the source has evolved into the HIMS instead of soft state. 

In Fig. \ref{fig:2}, Swift J1753.5-0127 traces a clear hard line in RID at the beginning of observations (red stars), with expected highest absolute rms. By extending this line to lower intensity (black line), we found a clear turn below $\sim$ 100 cts/s/PCU2, from where the RID split into two tracks. One track mostly follows the hard line, the other goes to the right side of the line. After applying energy spectra analysis, we found that the photon index ($\Gamma$) of most outsiders are higher than 1.7 (see black dots in Fig. \ref{fig:6}), with hardness less than 0.6. Meanwhile, their fractional rms experienced a sudden drop during that time (second panel in Fig. \ref{fig:4}, dash line interval). Following \citet{mun11, sol13}, we therefore identify those observations as HIMS (black dot) in Fig. \ref{fig:2}. 

\subsection{Spectral-temporal correlations}

In Fig. \ref{fig:4}, we plot the hardness ratio, total fractional rms ($rms_{\rm{total}}$), rms of the two broad noise components ($L_1$ and $L_2$) and photon index ($\Gamma$) as a function of time. The source experienced several small risings during the whole observation, in which there were obvious spectra softenings (increase of $\Gamma$ and decrease of total rms). When the hardness ratio drops from $\sim 0.7$ to $\sim 0.35$ near MJD 55337, the corresponding ($\Gamma$) significantly increases from $\sim 1.7$  to $\sim 2.1$. Meanwhile, the total rms decreases from $\sim 20\%$ to $\sim 12\%$, all suggesting a transition to HIMS. However, the source failed to evolve to the soft state, considering the remaining high total rms ($\sim 12\%$). 

In Fig. \ref{fig:4}, except the decaying stage of the outburst, one can see that the total rms ($rms_{\rm{total}}$) changes consistently with the hardnes ratio and contrarily to the photon index $\Gamma$, characterizing a decrease in Comptonized emission during spectral softenings ($\Gamma$ increase). During the LHS, in the cases of QPO observations at the beginning of outburst, the rms of flat-top noise ($L_1$, red cross) slightly decreases with the hardness, while the rms of high frequency noise ($L_2$, blue cross) increases with hardness. For the rest of the observations in LHS, the PDS can be well explained by two noise components, except that a hump appeared when the source transitioned to HIMS. Both $L_1$ and $L_2$ evolve similar to $rms_{\rm{total}}$ in the plots. However, when the source transitioned into HIMS, $L_2$ became undetectable in some observations, and the PDS can be described by a single zero-centered Lorentzian. Therefore, we excluded those observations in the $rms_{i}$ - time correlation (see the 3rd panel in Fig. \ref{fig:4}).

\begin{figure*}
	\includegraphics[scale=.45]{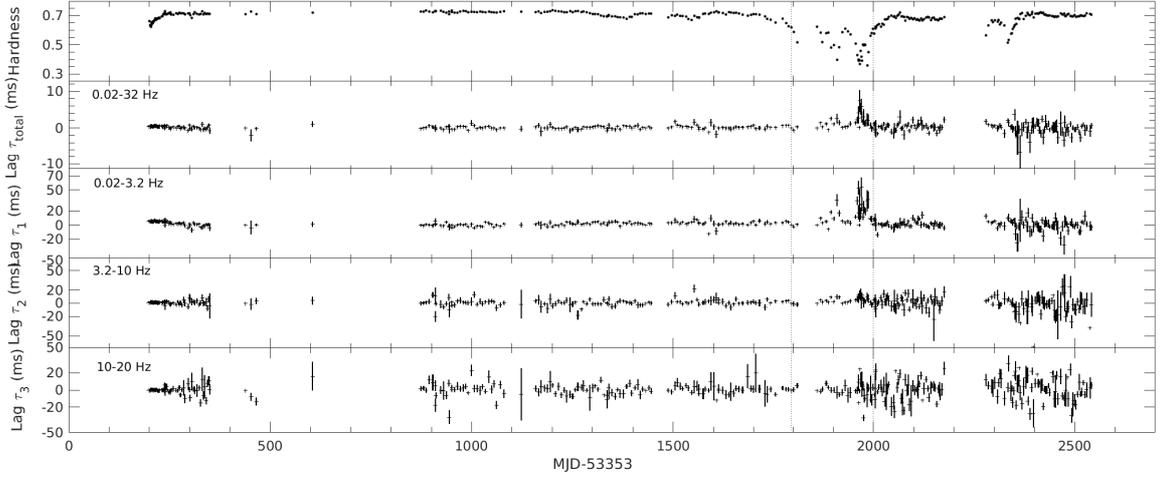}
	\caption{Hardness and X-ray time lags as a function of time. The lags are calculated between $\sim$ 2 -- 6 keV and $\sim$ 6 -- 15 keV. From the top to bottom, the X-ray time lags are averaged in the 0.02 -- 32 Hz, 0.02 -- 3.2 Hz, 3.2 -- 10 Hz and 10 -- 20 Hz frequency bands, separately. The HIMS interval is indicated by dash lines. \label{fig:5}}
\end{figure*}

In Fig. \ref{fig:5}, we plot the X-ray time lags averaged between $\sim$ 2 -- 6 keV and $\sim$ 6 -- 15 keV as a function of time. The plots show that the averaged total time lag (0.02 -- 32 Hz, $\tau_{\rm{total}}$, 2nd panel) increases significantly when spectra soften. Motivated by \citet{pot00}, we divided the frequency band into several sub-bands, 0.02 -- 3.2 Hz, 3.2 -- 10 Hz and 10 -- 20 Hz and computed the time lags in the same energy bands. The results suggest that the time lag averaged in 0.02 -- 3.2 Hz ($\tau_1$) contributed the most for the rise in the total time lag, while 3.2 -- 10 Hz ($\tau_2$) and 10 -- 20 Hz ($\tau_3$) time lags show a roughly constant evolution with time.

\begin{figure}
	\centering
	\includegraphics[scale=.45]{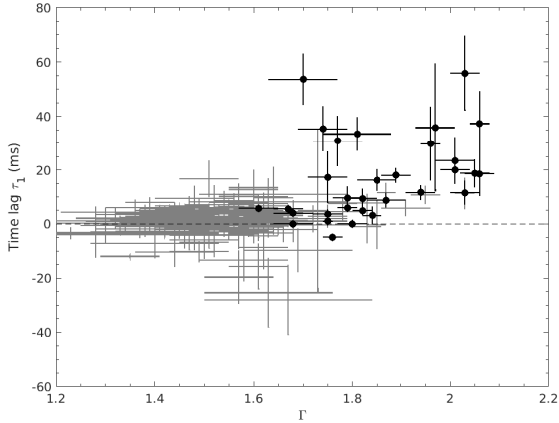}
	\caption{X-ray time lag in 0.02 -- 3.2 Hz frequency band  as a functions of photon index. The LHS and HIMS are plotted as grey and black dots separately. \label{fig:6}}
\end{figure}

To give a better illustration, we further plot the $\tau_1$ (0.02 -- 3.2 Hz) as a function of $\Gamma$ in Fig. \ref{fig:6}. It is very interesting to find that almost only hard lags are found when the source transited to HIMS (black dots), while zero lags are found during LHS. 

\begin{figure}
	\centering
	\includegraphics[scale=.4]{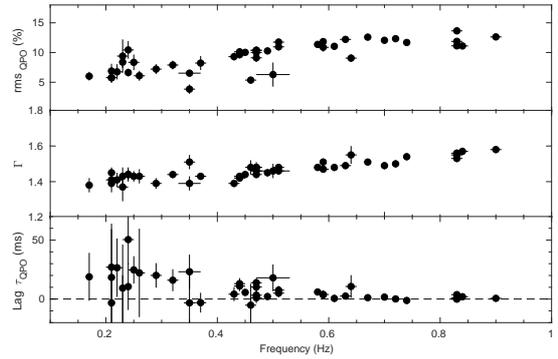}
	\caption{QPO rms, photon index and time lag as a function of QPO frequency (from top to bottom). \label{fig:7}}
\end{figure}
\begin{figure}
	\centering
	\includegraphics[scale=.43]{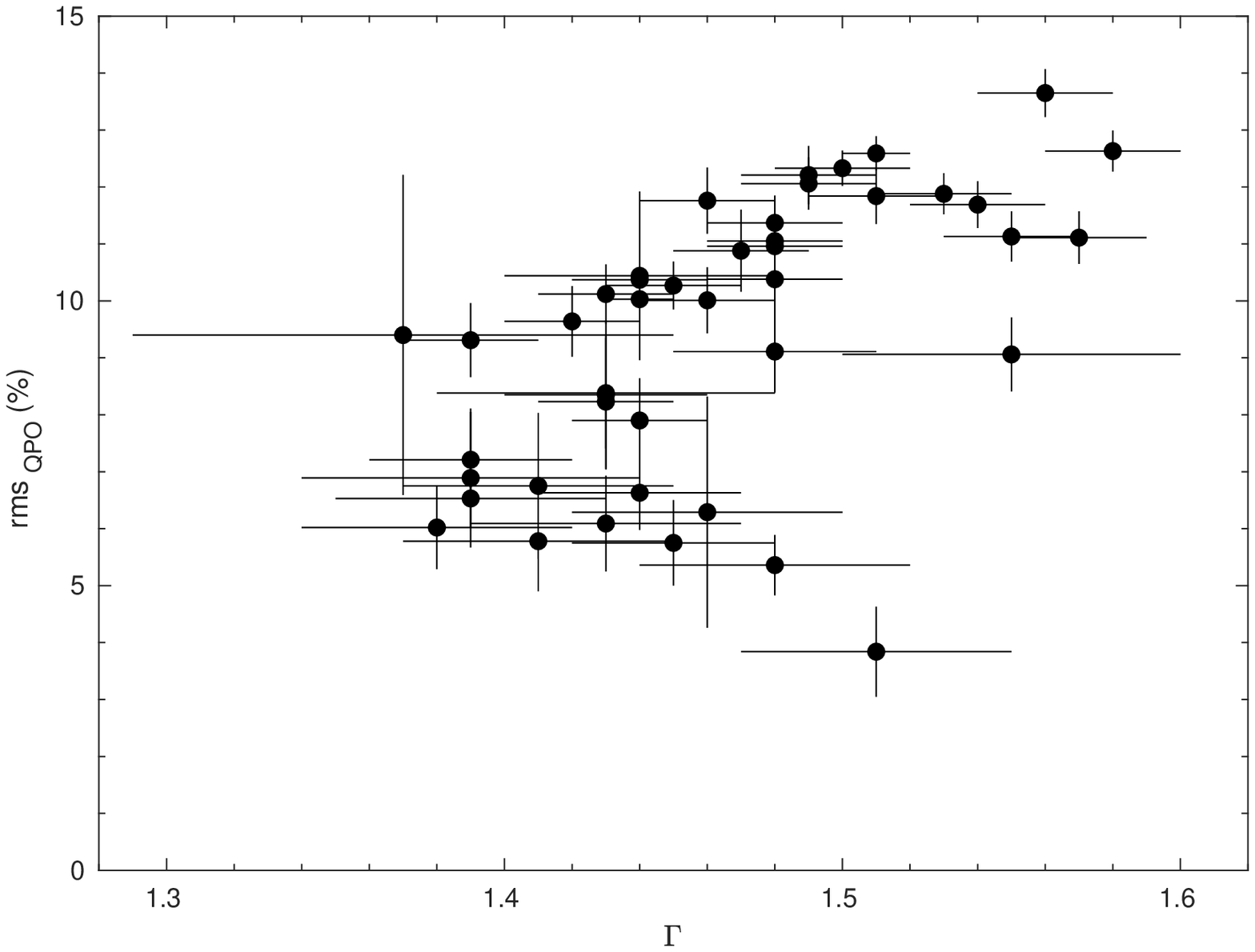}
	\caption{QPO fractional rms as a function of photon index. \label{fig:8}}
\end{figure}

Among the 344 observations, we found 46 LFQPOs, with most of which were detected at the beginning of the outburst, identifying as red stars in the figures. All the LFQPOs were recognized as type-C QPOs, given by the accompany of the flat-top noise and the corresponding source state. The QPOs are all detected in the low hard state, with the centroid frequency ranging from 0.1 to 0.9 Hz. In Fig. \ref{fig:7}, we plot the QPO fractional rms ($rms_{\rm{QPO}}$), photon index and time lag $\tau_{\rm{QPO}}$ as a function of QPO centroid frequency ($\nu_{\rm{QPO}}$). 

The QPO rms ($rms_{\rm{QPO}}$) and photon index ($\Gamma$) both positively increase with $\nu_{\rm{QPO}}$, from $\sim 5\%$ to $\sim 12\%$ and from $\sim 1.4$ to $\sim 1.6$. The QPO time lag generally shows a hard lag and decreases with $\nu_{\rm{QPO}}$ and $\Gamma$, from $\sim$ 50 ms to $\sim$ 0, despite of the errors. To investigate how QPO rms changes with spectral hardness, in Fig. \ref{fig:8}, we plot the QPO rms as a function of $\Gamma$. Despite the errors, the QPO rms seems to increase with $\Gamma$.

\section{Discussion} \label{sec:dis}

Using archival \textit{RXTE} data, we present a long-term spectral-temporal correlations of the black hole candidate Swift J1753.5-0127 from July 2, 2005 until November 29, 2011. In this work, we focused on the fast variability properties (rms, time lags, photon index) during its state transitions, and found that both the RID and time lag averaged in 0.02 -- 3.2 Hz can be used as an indicator to trace the state changes in this source. Unlike most BHTs, type-C QPOs were only found in the low hard state of Swift J1753.5-0127, while their centroid frequencies increase when spectra soften.

The HID and RID are widely used in BHTs to study the spectral evolution, in which both the hardness ratio and fractional rms are independent of spectral models. Similar to GX 339-4 \citep{mun11}, Swift J1753.5-0127 traces a hard line in RID during the LHS, with expected highest absolute rms, while it shows a clear turn during the HIMS, accompanied by lower absolute rms (Fig. \ref{fig:2}). Moreover, like most of the BHTs, the total rms is correlated with hardness ratio during HIMS (top and 2nd panels in Fig. \ref{fig:4}) \citep{bel11}. Although we observed several spectral hardenings and softenings in this source, we see that these observations follow a precise path in the RID. These results support the idea that the value of hardness ratio and the position in the RID are the indicators of spectral transitions to the HIMS. 

Similarities between Swift J1753.5-0127 and Cyg X-1 have been addressed in many works, both of them show frequency spectral softenings and hardenings during the ``failed transitions" and are unable to observe type-C QPOs in the HIMS. A similar correlation between photon indices of broken power law has been reported for both sources \citep{sol13, pot00}. In Section 3.2, we studied the time lag evolution in terms of several frequency bands, and found that the time lag increase significantly in the 0.02 -- 3.2 Hz band during HIMS. \textit{Besides, almost always hard lags are found during HIMS and increase with photon index $\Gamma$.} The origin of time lag in the BHTs are not yet well understood, it is widely believed that the time lag is scaled with the length of the region which gives birth to the observed photons. Hard lag during the HIMS suggests that the geometry of the disk-corona changes and some part of the source becomes larger in HIMS. \citet{now02} suggested an enlarging disk radius with a softening spectrum for a jet-like model to explain the increasing time lag in GX 339-4, in which the time lags generated from the propagation along the jet. 

Compare the evolution of PDS (Fig. \ref{fig:3}) and spectral fitting parameters together, we found that the whole PDS shifted to higher frequency when spectra softened. Meanwhile, the time lag increases in 0.02 -- 3.2 Hz band. The 0.02 -- 3.2 Hz band is dominated by the flat-top noise $L_1$ in the PDS. When the source transitioned to HIMS, the high frequency noise $L_2$ was suppressed by the occasional appearance of the $L_h$ and became undetectable when $\Gamma > 2.0$. Although no type-C QPOs has been found in the HIMS, $L_h$ was found processing similar properties with type-C QPOs compared to other BHTs \citep{sol13}, suggesting the black hole nature of the compact object. Besides, the type-C QPOs found in this source, with sometimes accompany of second harmonics, featuring the frequency modulation found in \citet{bu19}. A decreasing time lag with a increasing centroid frequency for type-C QPOs has been reported in GRS 1915+105 when frequency $< 2$ Hz \citep{zha17}. They suggested that the QPO time lag is dominated by the Compton effect in the corona and can be explained by a truncated accretion disk model. While the type-C QPOs origins from the Lense-Thirring precession of a hot inner flow within a truncated disk, the flat-top noise arises from the variations of mass accretion rate propagating inwards from the outer region of the accretion flow \citep{ing09, ing10}. During the LHS, when the accretion disk move inwards, the disk temperature and QPO frequency increase, the region of corona decreases, causing a decreasing time lag. This could also explain the increase hard lag in the 0.02 -- 3.2 Hz bands, since the photons from the inner region of the accretion flow is softer than the ones from outer region of the accretion flow. 

However, detailed spectral analysis had been performed in previous work, and suggested a cool inner disc extending near or close to the ISCO of Swift J1753.5-0127 \citep{hie09, mos13}. In this geometry, the type-C QPOs can either generate from the accretion disc or the hot corona. In principle, the hard lags can be caused by the propagating fluctuations on the accretion disc. In this case, the timescale of the hard lag should be in order of seconds \citep{kot01}, which is much larger than the case we found (less than 50 ms). On the other hand, the hard lags could be caused by the Compton effect in the corona, which will generate time lags in order of ms. In this scenario, the type-C QPOs are probably originated from the corona, which is suggested in \citet{bel11}. The QPO rms reaches its maximum fractional rms around $\Gamma = 1.55$ during LHS, implying there might be more than one component contributing to the QPO variability. However, more detailed study is needed to test this hypothesis.

\section{Conclusion} \label{sec:con}

We studied the evolution of outburst, the best fit PDS parameters, the X-ray time lags, the spectral parameters of the black hole candidate Swift J1753.5-0127 with \textit{RXTE} for more than 6 years. The source spent most of time in LHS, with frequency hardenings and softenings, and transitioned to HIMS near MJD 55337. We systematically study the long-term spectral-temporal correlated properties in a sample of 344 observations. 

Swift J1753.5-0127 traces a clear hard line in RID during the LHS, with expected highest absolute rms, while shows a clear turn during the HIMS, accompanied by lower absolute rms. Like most of the BHTs, the total rms is correlated with the hardness ratio during HIMS. These results support the idea that the value of hardness ratio and the position in the RID can be used an indicators to trace the state transitions in Swift J1753.5-0127. 

Different from Cyg X-1, we found that frequency-dependent time lag increased significantly in the 0.02 -- 3.2 Hz band during state transitions. The 0.02 -- 3.2 Hz time lag becomes positive and increases with photon index when the source turns into HIMS, which suggested that 0.02 -- 3.2 Hz time lag can be used as an indicator of state transition. QPO frequency correlates with its fractional rms and X-ray photon index and anti-correlates with the QPO time lag, featuring a moving forwards disc/corona model frame. 

\section*{Acknowledgements}

We thank the anonymous referee for the useful suggestions they made for our manuscript. This research has made use of data obtained from the High Energy Astrophysics Science Achieve Research Center (HEASARC) provided by NASA Goddard Space Flight Center. This work is supported by the National Program on Key Research and Development Project (Grant No. 2016YFA0400800), the National Natural Science Foundation of China (Grant No. U1838108, 11733009, 11673023, U1838110, U1838113, U1838111, U1838115 and U1838201), the CAS Pioneer Hundred Talent Program (Grant No. Y8291130K2) and the Scientific and technological innovation project of IHEP (Grant No. Y7515570U1).

\clearpage
\section*{Appendix}
Here we report four tables (Table 1 - 4) in which we show the averaged fractional rms, time lags, QPO parameters, as well as the best-fitting parameters of energy spectra for all the observations.
\begin{center}
\begin{deluxetable}{ccccc}
		\tablecaption{Integrated fractional rms computed from the power density spectra extracted from the \textit{RXTE}/PCA observations of Swift J1753.5-0127. The details on the extraction of the PDS and on the fitting procedures are described in Section 2.2. For each observation, the table shows the total fractional rms (rms$_{\rm{total}}$), rms of $L_1$ (rms$_1$) and $L_2$ (rms$_2$) with $1\sigma$ errors. \label{tab:1}}
		\tablehead{
			\colhead{Obs. ID} & \colhead{MJD} & \colhead{rms$_{\rm{total}}$ (\%)} & \colhead{rms$_1$ (\%)} & \colhead{rms$_2$ (\%)}
		}
		\startdata
		91094-01-01-00 &  53553.1 &   25.22$\pm$0.21 &    13.96$\pm$0.69 &    11.16$\pm$0.49 \\
		91094-01-01-01 &  53555.2 &   26.24$\pm$0.14 &    14.40$\pm$0.42 &    11.54$\pm$0.21 \\
		91094-01-01-02 &  53555.6 &   26.49$\pm$0.14 &    14.60$\pm$0.41 &    11.62$\pm$0.22 \\
		91094-01-01-03 &  53556.2 &   26.52$\pm$0.14 &    14.39$\pm$0.36 &    12.07$\pm$0.16 \\
		91423-01-01-04 &  53557.2 &   26.36$\pm$0.13 &    13.54$\pm$0.32 &    11.79$\pm$0.19 \\
		91094-01-01-04 &  53557.5 &   26.86$\pm$0.13 &    13.77$\pm$0.35 &    12.05$\pm$0.15 \\
		91423-01-01-00 &  53558.3 &   26.91$\pm$0.12 &    12.81$\pm$0.35 &    11.82$\pm$0.14 \\
		91094-01-02-01 &  53559.7 &   26.47$\pm$0.15 &    13.16$\pm$0.48 &    11.87$\pm$0.21 \\
\enddata
		\tablecomments{The noise components were fitted with zero-centered Lorentzian function in our work. The fit parameters of QPOs are given in Table 2.}
	\end{deluxetable}
	
	\clearpage
	\begin{deluxetable}{cccccc}
		\tablecaption{Fit parameters of the 46 LFQPOs observations of Swift J1753.5-0127 extracted from \textit{RXTE}. The table shows the centroid frequency ($\nu_{\rm{QPO}}$), the quality factor Q, fractional rms (rms$_{\rm{QPO}}$) and the time lag ($\tau_{\rm{QPO}}$) of QPOs. The details on the extraction of the spectra and on the fitting procedures are described in Section 2.2. \label{tab:2}}
		\tablehead{
			\colhead{Obs. ID} & \colhead{MJD} & \colhead{$\nu_{\rm{QPO}}$ (Hz)} & \colhead{Q} & \colhead{rms$_{\rm{QPO}}$ (\%)} & \colhead{$\tau_{\rm{QPO}}$ (ms)}}
		\startdata
		91094-01-01-00 &  53553.1 &    0.64$\pm$0.01 &     4.64$\pm$0.79 &     9.06$\pm$0.65 &    10.67$\pm$9.62 \\
		91094-01-01-01 &  53555.2 &    0.83$\pm$0.01 &     3.97$\pm$0.42 &    11.13$\pm$0.44 &     3.75$\pm$2.01 \\
		91094-01-01-02 &  53555.6 &    0.84$\pm$0.01 &     4.58$\pm$0.58 &    11.11$\pm$0.46 &     2.00$\pm$2.44 \\
		91094-01-01-03 &  53556.2 &    0.83$\pm$0.01 &     4.39$\pm$0.33 &    11.88$\pm$0.36 &    -0.08$\pm$1.68 \\
		91423-01-01-04 &  53557.2 &    0.90$\pm$0.01 &     3.71$\pm$0.28 &    12.63$\pm$0.36 &     0.61$\pm$1.59  \\
		91094-01-01-04 &  53557.5 &    0.83$\pm$0.01 &     2.98$\pm$0.23 &    13.65$\pm$0.42 &     1.12$\pm$1.41 \\
		91423-01-01-00 &  53558.3 &    0.67$\pm$0.00 &     4.06$\pm$0.27 &    12.59$\pm$0.30 &     1.18$\pm$1.51 \\
		91094-01-02-01 &  53559.7 &    0.74$\pm$0.00 &     4.91$\pm$0.47 &    11.69$\pm$0.41 &    -1.21$\pm$2.74 \\
\enddata
		\tablecomments{Q is the quality factor, which is defined as $Q=\nu_{\rm{QPO}}/\rm{FWHM}$, as indication of the coherence of a QPO signal. QPO time lags were computed between $\sim$ 2 -- 6 keV and $\sim$ 6 -- 15 keV. }
	\end{deluxetable}

	\clearpage
	\begin{deluxetable}{cccccc}
		\tablecaption{X-ray time lags of Swift J1753.5-0127 extracted from \textit{RXTE}, averaged over the 0.02 -- 32 Hz ($\tau_{\rm{total}}$), 0.02 -- 3.2 Hz ($\tau_1$), 3.2 -- 10 Hz ($\tau_2$) and 10 -- 20 Hz ($\tau_3$) frequency bands separately. The details on the extraction of the cross spectra are described in Section 2.2. \label{tab:3}}
		\tablehead{
			\colhead{Obs. ID} & \colhead{MJD} & \colhead{$\tau_{\rm{total}}$ (ms)} & \colhead{$\tau_1$ (ms)} & \colhead{$\tau_2$ (ms)} & \colhead{$\tau_3$ (ms)}}
		\startdata	
		91094-01-01-00 &  53553.1 &    0.41$\pm$0.18 &     5.26$\pm$1.52 &     1.24$\pm$1.45 &    -0.58$\pm$1.65 \\
		91094-01-01-01 &  53555.2 &    0.68$\pm$0.05 &     6.65$\pm$0.44 &     2.44$\pm$0.34 &     0.69$\pm$0.37 \\
		91094-01-01-02 &  53555.6 &    0.56$\pm$0.05 &     6.10$\pm$0.48 &     2.01$\pm$0.37 &    -0.11$\pm$0.41 \\
		91094-01-01-03 &  53556.2 &    0.58$\pm$0.04 &     5.53$\pm$0.37 &     2.25$\pm$0.26 &     0.99$\pm$0.28 \\
		91423-01-01-04 &  53557.2 &    0.64$\pm$0.04 &     5.57$\pm$0.42 &     2.52$\pm$0.30 &     1.34$\pm$0.36 \\
		91094-01-01-04 &  53557.5 &    0.72$\pm$0.04 &     6.57$\pm$0.38 &     2.78$\pm$0.26 &     1.28$\pm$0.27 \\
		91423-01-01-00 &  53558.3 &    0.60$\pm$0.03 &     6.24$\pm$0.29 &     1.55$\pm$0.24 &     0.75$\pm$0.24 \\
		91094-01-02-01 &  53559.7 &    0.65$\pm$0.05 &     6.03$\pm$0.52 &     2.22$\pm$0.39 &     0.97$\pm$0.44 \\
\enddata
	\end{deluxetable}
	\clearpage
	\begin{deluxetable}{cccccc}
		\tablecaption{Best-fitting parameters of 344 energy spectra of Swift J1753.5-0127 extracted from RXTE observatimeons performed
			during the whole time. The details on the extraction of the spectra and on the fitting procedures are described in Section
			2.3. For all the observations performed before 2009 December we made use of PCA and HEXTE data while for the remaining
			observations only PCA data were available. Here, we report the values of the parameters used to fit
			the spectra, as well as the unabsorbed fluxes in the 3 -- 25 keV energy range. \label{tab:4}}
		\tablehead{ \colhead{Obs. ID} & \colhead{$\chi^2$/d.o.f} & \colhead{flux\tablenotemark{a} ($\rm{erg/s/cm^2}$)} & \colhead{$\Gamma$} & \colhead{kT (keV)} & \colhead{$R_{in}\tablenotemark{b}\sqrt(\cos(\theta))$ (km)}}
		\startdata
		PCA+HEXTE \\
		91094-01-01-00  &    83.31/87 &    44.33 &     1.55$\pm$0.05 &     1.37$\pm$0.15 &     2.01$\pm$0.28 \\
		91094-01-01-01  &   120.75/87 &    66.08 &     1.55$\pm$0.02 &     1.35$\pm$0.05 &     3.04$\pm$0.16 \\
		91094-01-01-02  &    73.10/87 &    68.57 &     1.57$\pm$0.02 &     1.35$\pm$0.05 &     3.08$\pm$0.17 \\
		91094-01-01-03  &   107.74/87 &    69.79 &     1.53$\pm$0.02 &     1.38$\pm$0.04 &     3.08$\pm$0.14 \\
		91094-01-01-04  &    92.04/87 &    74.07 &     1.56$\pm$0.02 &     1.35$\pm$0.05 &     3.27$\pm$0.17 \\
		91094-01-02-00  &    78.49/87 &    72.22 &     1.50$\pm$0.02 &     1.43$\pm$0.04 &     2.92$\pm$0.12 \\
		91094-01-02-01  &    86.18/87 &    72.47 &     1.54$\pm$0.02 &     1.40$\pm$0.05 &     2.86$\pm$0.16 \\
\enddata
		\tablecomments{a: The value of the flux is in units of $10^{-10} \rm{erg/s/cm^2}$. b: The inner disc radius can be derived from the normalization N of the \textit{diskbb} component: $R_{in}=D_{10}*\sqrt(N)/\sqrt(\cos(\theta))$, where $D_{10}$ is the source distance in units of 10 kpc and $\theta$ is the disc inclination. We used D= 8 kpc to calculate the $R_{in}$.}
	\end{deluxetable}
\end{center}
\bsp	
\label{lastpage}

\begin{thebibliography}{}
	
	\bibitem[Belloni \& Hasinger(1990)]{bel90} Belloni, T. \& Hasinger, G.\ 1990, A\&A, 230, 103
	\bibitem[Belloni et al.(2011)]{bel11} Belloni T., Motta S., Mu\~noz-Darias T.\ 2011, Bull. Astron. Soc. India, Vol. 39, No. 3, p. 409 
	\bibitem[Belloni \& Motta(2016)]{bel16} Belloni, T. M., \& Motta, S. E.\ 2016, Transient black hole binaries, In Astrophysics of Black Holes (pp. 61-97), Springer, Berlin, Heidelberg
	\bibitem[Bu et al.(2019)]{bu19} Bu, Q., Belloni, T, Chen, L., Li, Z., Qu, J.\ 2019, Under Review
	\bibitem[Casella et al.(2005)]{cas05} Casella, P., Belloni, T., \& Stella, L.\ 2005, \apj, 629, 403
	\bibitem[Capitanio et al.(2009)]{cap09} Capitanio F., Belloni T., Del Santo M., Ubertini P.\ 2009, \mnras, 398, 1194
	\bibitem[Chatterjee et al.(2016)]{cha16} Chatterjee, D., Debnath, D., Chakrabarti, S. K., et al.\ 2016, \apj, 827, 88
	\bibitem[Debnath et al.(2015)]{deb15} Debnath, D., Molla, A. A., Chakrabarti, S. K., et al.\ 2015, \apj, 803, 59
	\bibitem[Done et al.(2007)]{don07} Done, C., Gierlinski, M., \& Kubota, A.\ 2007, A\&ARv, 15, 1
	\bibitem[Froning et al.(2014)]{fro14} Froning C. S., Maccarone T. J., France K., et al.\ 2014, \apj, 780, 48
	\bibitem[Hiemstra et al.(2009)]{hie09} Hiemstra, B., Soleri, P., Mendez, M., et al.\ 2009, \mnras, 394, 2080
	\bibitem[Homan et al.(2001)]{hom01} Homan, J., Wijnands, R., van der Klis, M., et al.\ 2001, \apjs, 132, 377
	\bibitem[Mostafa et al.(2013)]{mos13} Mostafa, R., Mendez, M., Hiemstra, B. et al.\ 2013, \mnras, 431, 2431
	\bibitem[Ingram et al.(2009)]{ing09} Ingram, A., Done, C., \& Fragile, P. C.\ 2009, \mnras, 696, 1257
	\bibitem[Ingram et al.(2010)]{ing10} Ingram, A., Done, C.\ 2010, \mnras, 405, 2447
	\bibitem[Jana et al.(2016)]{jan16} Jana, A., Debnath, D., Chakrabarti, S. K., et al.\ 2016, \apj, 803, 107  
	\bibitem[Kotov et a.(2001)]{kot01} Kotov, O., Churazov, M., Gilfanov, M.\ 2001, \mnras, 327, 799       
	\bibitem[Leahy et al.(1983)]{lea83} Leahy D. A., Elsner R. F., Weisskopf M. C.\ 1983, \apj, 272, 256
	\bibitem[Motta et al.(2015)]{mot15} Motta, S., Casella P., Henze, M., et al.\ 2015, \mnras, 447, 2059
	\bibitem[Motta et al.(2017)]{mot17} Motta, S., Rouco-Escorial, A., Kuulkers, E., et. al.\ 2017, \mnras, 468, 2311
	\bibitem[Mu\~noz-Darias et al.(2011)]{mun11} Mu\~noz-Darias, T., Motta S., Belloni T. M.\ 2011, \mnras, 410, 679
	\bibitem[Mu\~noz-Darias et al.(2014)]{mun14} Mu\~noz-Darias, T., Fender, R., Motta, S., \& Belloni, T.\ 2014, \mnras, 443, 3270
	\bibitem[Nowak et al.(2002)]{now02} Nowak, M. A., Wilms, J., \& Dove, J. B.\ 2002, \mnras, 332, 856
	\bibitem[Palmer et al.(2005)]{pal05} Palmer, D. M., Barthelmey, S. D., Cummings, J. R., et al.\ 2005, ATel, 546, 1
	\bibitem[Pottschmidt et al.(2000)]{pot00} Pottschmidt, K., Wilms, J., Nowak, M. A., et al.\ 2000, A\&A, 357, L17  
	\bibitem[Pottschmidt et al.(2003)]{pot03} Pottschmidt, K., Wilms, J., Nowak, M. A., et al.\ 2003, A\&A, 407, 1039 
	\bibitem[Protassov et al.(2002)]{pro02} Protassov R., van Dyk D. A., Connors A.,et al.\ 2002, \apj, 571, 545
	\bibitem[Remillard et al.(2002)]{rem02} Remillard, Ronald A et al.\ 2002, \apj, 564, 962
	\bibitem[Shaw et al.(2016)]{sha16} Shaw, A. W., Gandhi, P., Altamirano, D., et al.\ 2016, \mnras, 458, 1636
	\bibitem[Stella \& Vietri(1999)]{ste99} Stella, L., Vietri, M. \& Morsink, S. M.\ 1999, \apj, 524, 63
	\bibitem[Soleri et al.(2013)]{sol13} Soleri, P., Mu\~noz-Darias, T., Motta, S., et al.\ 2013, \mnras, 429, 1244  
	\bibitem[Tagger \& Pellat(1999)]{tag99} Tagger M., Pellat R.\ 1999, A\&A, 349, 1003
	\bibitem[Titarchuk \& Osherovich(1999)]{tit99} Titarchuk L., Osherovich V.\ 1999, \apj, 518, L95
	\bibitem[van den Eijnden et al.(2016)]{van16} van den Eijnden, J., Ingram, A., Uttley, P., et al.\ 2016, \mnras, 464, 2643
	\bibitem[Wijnands \& van der Klis(1999)]{wij99} Wijnands, R. \& van der Klis, M.\ 1999a, \apj, 514, 939
	\bibitem[Yoshikawa et al.(2015)]{yos15} Yoshikawa A., Yamada S., Nakahira S., et al.\ 2015, \pasj, 67, 11
	\bibitem[Zhang et al.(1995)]{zha95} Zhang, W., Jahoda, K., Swank, H., et al.\ 1995, \apj, 449, 930 
	\bibitem[Zhang et al.(2017)]{zha17} Zhang, L., Wang, Y., Mendez, M., et al.\ 2017, \apj, 845, 143
	
\end{thebibliography}
\end{document}